# Theoretical Investigation of Size, Shape and Aspect Ratio Effect on the LSPR Sensitivity of Hollow-Gold Nanoshells

M. Shabaninezhad[a1], and Guda Ramakrishna[b]

[a]Department of Physics, Western Michigan University, Kalamazoo MI 49008

[b]Department of Chemistry, Western Michigan University, Kalamazoo MI 49008

**KEYWORDS:** *LSPR, DDA, Multilayer Mie Theory, Plasmon sensitivity*

**ABSTRACT:** The change in refractive index around plasmonic nanoparticles upon binding to biomolecules is routinely used in localized surface plasmon resonance (LSPR) based biosensors and in bio-sensing platforms. In this study, the plasmon sensitivity of hollow gold (Au) nanoshells is studied using theoretical modeling where the influence of shape, size, shell thickness and aspect ratio are addressed. Different shapes of hollow Au nanoshells are studied that include: sphere, disk, triangular prism, rod, ellipsoid, and rectangular block. Multi-layered Mie theory and discrete dipole approximation (DDA) were used to determine the LSPR peak position, and LSPR sensitivity as a function of size, shell thickness, shape, and aspect ratio. The change in LSPR peak wavelength per unit refractive index is defined as the sensitivity, and interesting results were obtained from the analysis. The rectangular block and rod-shaped Au nanoshells have shown maximum LSPR sensitivity when compared to other shaped Au nanoshells. In addition, increased sensitivity was observed for higher aspect ratio as well as for smaller shell thicknesses. The results are rationalized based on the inner and outer surface plasmonic coupling.

**I. INTRODUCTION**

Plasmonic sensors have attracted enormous scientific interest in recent years as the optical properties are governed by strong coupling between the incident light and conduction electrons leading to LSPR.[1–27] The LSPR leads to strong scattering and absorption of incident light, resulting in highly localized and amplified electric field around the NPs[24–26,28–32] and plays a vital role in applications of plasmonic NPs.[1–27] Among plasmonic NPs, gold NPs have received significant attention for biological and chemical applications as they are non-toxic, biocompatible, inert and ease of synthesis.[33,34] Also, Au nanoparticles possess strong extinction spectra with LSPR peak that can be tuned from visible to near infrared region based on their size, shape and configuration.[14] The tunability of LSPR peak into the near infrared region is highly desirable, especially for biological applications as the human tissue has the highest transparency in this frequency region.[25,35] The LSPR maximum of plasmonic NPs is very sensitive to the refractive index (RI) of the surrounding medium and shifts to longer wavelengths with an increase in RI. Since the RI of biological or chemical analytes is usually higher than the RI of the surrounding medium,[36] binding of the molecules to the surface of the plasmonic NPs alters the effective RI of the surrounding medium, redshifts the LSPR maximum and changes the electric field intensity around the NP.[1–27,37–43] In addition, the localized enhanced electric field around plasmonic NPs decay exponentially with increasing distance from the surface that creates a small sensing volume and detects the target molecules in the close vicinity of the NPs.[9] The LSPR peak shift, which is the basis of the majority of LSPR sensing applications, can be detected by measuring scattered light from the NPs.[9] However, to sense the RI change induced LSPR peak shift of the small-sized NPs that weakly scatter incident light, ensemble extinction methods are commonly used.[9]

---

[1] Corresponding Author: masoud.shabaninezhadnavrood@wmich.edu

LSPR biosensors offer a wide range of advantages such as low-cost, higher surface to volume ratio, facile surface chemistry, detecting of several different targets at the same time and monitoring binding of molecules in real-time.[3,4,14,26] For example, a gold nanorod with different aspect ratios have been used for multiplex sensing by monitoring the distinct LSPR peak shift caused by the binding of three different molecules on their surfaces.[26] In another study, to determine the concentration of the microRNAs (miRs) in the plasma or blood of pancreatic cancer patients, Joshi et al. applied nanoprism Au nanoparticles to detect and measure miRs.[42] Using LSPR sensing technique, they observed that concentration of miR-10b is almost four times greater than of miR-21 in pancreatic cancer patients.[42] More recently, Sriram et al. developed an analyzing technique to detect interleukin-6 (IL-6) protein based on the changing in color from plasmonic gold nanoparticles using dark field microscopy and digital camera with a CMOS sensor.[44] They were able to detect 4.76 nM (100 ng/mL) of IL-6 using this technique. However, the main drawback of this method is that its working wavelength range is from 450 nm to 625 nm and can be applied only for NPs that have single plasmon peak.

In contrast to the vast majority of the biological sensors, which need labels such as fluorophores, radioisotopes or enzyme to generate and enhance target signals,[7] LSPR based biosensors are label-free.[8,40,41] Due to enhanced electric field around NPs, the target molecule signals can be amplified and be used to detect the molecules without chemical or biological labels.[8,40,41] In addition, unlike surface plasmon resonance (SPR) sensing that has been widely used for detecting bimolecular interactions,[45] LSPR sensing is not sensitive to temperature and requires simple instrumentation.[46] At the same time, LSPR leads to enhanced localized electric field around plasmonic NPs that have shorter decay length in comparison to SPR and results in better sensitivity to changing RI in the close vicinity of the NPs.[9,25] However, despite the advantages mentioned above, the sensitivity and the signal-to-noise ratio of LSPR sensing is much less than the SPR sensing technique.[9,14] Thus, designing and optimizing different LSPR plasmonic nanostructures to obtain higher sensitivity by changing the refractive index of the surrounding medium is desirable for sensing applications.[14,27]

The LSPR sensitivity of plasmonic NPs is widely defined as LSPR peak shift per surrounding medium refractive index unit (RIU)[47]

$$S = \frac{d\lambda_{LSPR}}{dn} \qquad (1)$$

In addition to medium effect, the LSPR peak position and its shift, electric field enhancement and its decay length can be adjusted by changing parameters such as shape and size of the NPs, and the polarization direction of incident light.[5,13,25,29,31,37,38,48–54] All these optical properties can also be adjusted by varying inter-particle distances of neighboring nanoparticles in dimer or trimer structures[55–59] or varying shell thickness in core-shell nanostructures.[60,61] Several research efforts have focused on the shape, size and other parameter effects on the sensitivity of the plasmonic NPs.[2,7,8,11,38,39,53,62,63] For example, a sensitivity factor of 408.8 nm/RIU has been reported for a spherical hollow core- Au nanoshells with a mean total size of 50 nm and shell thickness of 4.5 nm, which was almost 6-times stronger than the solid Au NPs with the same diameter.[62] In this study, the core region was filled out with the surrounding medium.[62] In theoretical work, Jain et al. investigated LSPR sensitivity of spherical silica- gold core-shell with a total diameter of 80 nm and demonstrated that LSPR sensitivity of nanoshells versus shell-to-core ratio shows universal scaling behavior.[64] In addition, due to increased plasmonic coupling between inner and outer surface layers of shell by reducing shell thickness, they had observed significant enhancement of LSPR sensitivity from 129 nm/RIU to 363 nm/RIU when shell thickness decreased from 40 nm (solid sphere) down to 4 nm.[64] In another study, a nanorice nanoshell with longitudinal diameter of 340± 20 nm, transverse diameter of 54± 4 nm and shell thickness of 13.1± 1.1 nm has shown LSPR sensitivity of 801.4 and 103.0 nm/RIU for longitudinal and transverse plasmon modes, respectively.[53] Lee et al. have shown that the LSPR sensitivity of Au nanoparticles increases with increase in the size and the AR. Interestingly, they observed 491.4 (nm/RIU) LSPR sensitivity for Au nanorod with an aspect ratio of 3.4 and effective radius of 20 nm was almost 3.2 and 1.5 times higher when compared to spherical Au NPs with r=20 nm and r=60 nm, respectively.[25] In another study, by varying AR of gold nanorods (with an effective diameter of 50 nm) from 2 to 6, LSPR sensitivity in the range of 200-600 (nm/RIU) has been obtained from FDTD simulations.[65] Dondapati and coworkers have used biotin-modified gold nanostars to sense binding of streptavidin with concentrations as low as 0.1 nM by measuring the plasmon shift.[8] These nanoparticles have shown average sensitivity of 218 nm/RIU.[8] In a recent work, Woo et al. elucidated anisotropic property of arrays of Au nanodisks with a structural dimension of 162 and 340 nm, and measured sensitivity of 327 nm/RIU and 167 nm/RIU for longitudinal and transverse modes, respectively.[14]

Although extensive research has been carried out on LSPR sensitivity of the different shape solid plasmonic NPs[2,8,25,26,39–41,66] and spherical nanoshells,[37,53,62,64] there has been little quantitative analysis of size, shape, shell thickness and aspect ratio effects on the LSPR sensitivity of the hollow/Au nanoshells. In this paper, we focused on understanding these parameters effects on the plasmonic properties of hollow/Au nanoshells in order to obtain structures with higher sensitivity. The main objectives of this study are to probe the effect of (1) size and shell thickness, (2) shape and shell thickness, and (3) aspect ratio on LSPR sensitivity. Multi-layered Mie theory was used to calculate the size effect of LSPR sensitivity for different size gold sphere from 20 to 80 nm. Then, to understand the shape effect, we have performed simulations on disk, rod, ellipsoid, rectangular block, and prism shape NPs with an effective diameter of 40 nm by varying shell thicknesses (see FIG. 1). Also, we have studied the AR effect on plasmonic properties of rod and rectangular block shaped hollow/Au nanoshells while varying AR from 1 to 5 for a fixed effective diameter of 40 nm. Section II describes the calculation methods and results and discussion is provided in Section III.

## II. CALCULATION METHODS

The dielectric function of the bulk metal can be described by the classical Drude model.[67] In this model, metals are made of the free electron gas, which oscillates around positive ions. The electric response of the metal electrons in plasma to the external electric field as a function of incident light frequency is given by[67]

$$\varepsilon(\omega) = \varepsilon_{IB}(\omega) - \frac{\omega_p^2}{\omega^2 + i\Gamma_0\omega} \quad (2)$$

where $\omega$ is the frequency of the incident light, $\Gamma_0$ represents collision frequency of the electrons in bulk metal, and $\omega_p$ is plasma frequency that depends on electron density. $\varepsilon_{IB}$ denotes interband transition contribution of the transferred electrons from valence d-band to conduction band in dielectric function of the metal where is a function of incident light frequency. However, in the nanoscale regime, when the size of NPs is smaller than the mean free path of free electrons in bulk metal, electron- surface collisions will increase, leading to enhanced damping frequency. The modified collision frequency of the NPs can be described as:[67]

$$\Gamma(r_{eff}) = \Gamma_0 + \frac{AV_F}{r_{eff}} \quad (3)$$

here $r_{eff}$, $V_F$, and A are the effective radius of the shell NPs, Fermi velocity, and empirical fitting parameter, respectively. By considering this effect, the modified dielectric function of NPs is given by[59]

$$\varepsilon_{NP}(\omega, r_{eff}) = \varepsilon_{bulk}(\omega) + \frac{\omega_p^2}{\omega^2 + i\Gamma_0\omega} - \frac{\omega_p^2}{\omega^2 + i\Gamma(r_{eff})\omega} \quad (4)$$

In this work, for bulk Au, values of the dielectric function at different wavelengths were taken from experimental data provided by Johnson and Christy.[68] In addition, $\omega_p = 1.35 \times 10^{16}$ rads$^{-1}$,[68] $\Gamma_0 = 1.06 \times 10^{14}$ rads$^{-1}$,[29] $V_F = 1.4 \times 10^6$ ms$^{-1}$, and A=1 have been chosen in our simulation.

In the simulation, multi-layered Mie theory[69,70] and discrete dipole approximation (DDA)[71–74] have been used to obtain optical properties of the spherical and non-spherical nanoshells, respectively. All codes were developed by our group in MATLAB Software. To test the DDA codes, the results of DDA code for spherical nanoshells were compared against multi-layered Mie theory, and there was an excellent agreement in the calculation of extinction, absorption and scattering efficiencies by both methods. Then, to check our DDA codes for non-spherical hollow/Au nanoshells, we compared the results with available spectra of solid NPs in the literature[54] for three cases: 1) setting the refractive index of core to be equal to refractive index of the shell, 2) setting the effective radius of the core to be zero, and 3) vanishing shell thickness. In all of the cases, there was an excellent agreement between our results and available data in the literature.

FIG. 1 shows the schematic of six different (Sphere, Disk, Triangular Prism, Rod, Ellipsoid, and Rectangular block) hollow-Au core-shell structures that have been investigated in this research. Rod NPs are made of one cylinder and hemisphere caps at both ends. In addition, to obtain appropriate mesh size for the non-spherical nanoshells, $C_{ext}$ of samples was calculated with different mesh sizes (d=0.25 nm, d=0.35 nm, and d=0.5 nm) to show convergence in the simulation. Extinction spectra of Rod, ellipsoid, and rectangular block with shell thickness of t > 0.3 $r_{eff}$ have been converged at d=0.5 nm, while for t ≤ 0.3 $r_{eff}$, d= 0.25 nm has been employed to collecting data. For Disk and Prism hollow nanoshells, d has been set to 0.25 nm to collect the data. FIG. S1 shows an example of the convergence of extinction spectra of 40 nm ellipsoid hollow/Au nanoshells with a shell thickness of a) t=0.1 $r_{eff}$ and b) t=0.2 $r_{eff}$.

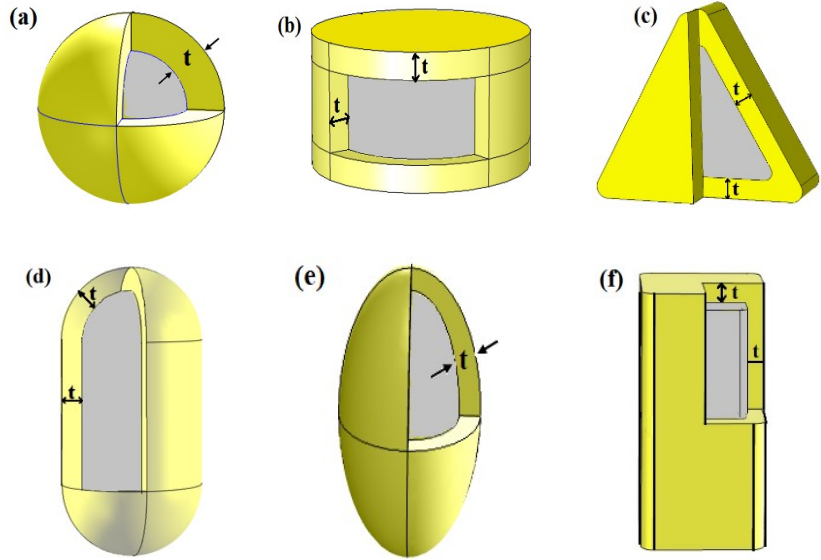

**FIG. 1.** Schematic of (a) Spherical, (b) Disk, (c) Triangular Prism, (d) Rod, (e) Ellipsoid and (f) Rectangular block, hollow/Au nanoshells that have been investigated in this work. Sharp corners of the rectangular block and prism have been rounded. The rounding radius for the outer surface set to be 2 nm and the inner radius has been calculated by multiplying ratio of the core-to-total size to 2 nm.

## III. RESULTS AND DISCUSSIONS

### A. Dielectric function effect on sensitivity and FWHM

In order to theoretically elucidate the shell thickness influence on the dielectric function and consequently on plasmon sensitivity and FWHM of the hollow/Au nanoshells, Quasi-static approximation was used. In this approximation, the size of the NPs is much smaller than incident light wavelength (kd<<1) whereby the electromagnetic phase shift, and electric field and polarizability variation are negligible inside the Np.[75] In this model, by considering dipole mode and ignoring higher order modes, the absorption spectrum inside Np is given by[75]

$$\sigma_{abs}(\omega) = \frac{V}{3c}\varepsilon_m^{1.5}\sum_{i=1}^{3}\frac{1}{L_i^2}\frac{\omega\varepsilon_2}{\left[\varepsilon_1+\varepsilon_m(\frac{1}{L_i}-1)\right]^2+\varepsilon_2^2} \quad (5)$$

where V, c and $\varepsilon_m$ are the volume of the nanoparticle, the speed of the light in a vacuum and the dielectric function of the surrounding medium. $L_i$ represent shape factors related to depolarization effect and are equal to 1/3, 1 and 0 for spherical, flat disk and infinite spheroid (when the incident light is in direction of the long axis).[75] In addition, $\varepsilon_1$ and $\varepsilon_2$ are the real and imaginary part of the dielectric function of the metallic nanoparticles. In order to obtain the corresponding wavelength of LSPR, the denominator of the Eq. 5 should be minimum that gives[75]

$$\varepsilon_1 + \varepsilon_m\left(\frac{1}{L_i}-1\right) = 0 \quad (6)$$

with minimizing the denominator of $\sigma_{abs}$ in Eq. 5, LSPR frequency can be expressed as[75]

$$\omega_{LSPR} = \frac{\omega_p}{\left[\varepsilon_{IB}+\varepsilon_m(\frac{1}{L_i}-1)\right]^{\frac{1}{2}}} \quad (7)$$

From the above equation, it is evident that $\omega_{LSPR}$ decreases with increasing the refractive index of the medium. This shift to longer wavelengths can be ascribed to decreasing the restoring Coulombic force due to increased shielding effect between the oscillating electrons and positive ions that results in a reduction of the required incident light energy to excite coherent oscillation of the electrons.[5]

To evaluate hollow/Au nanoshells electric permittivity, we used the composition-weighted average to calculate $\varepsilon_{eff}$ of the structure[25,76–78]

$$\varepsilon_{eff}(\omega) = X_{core}\,\varepsilon_{core}(\omega) + (1 - X_{core})\varepsilon_{shell}(\omega) \tag{8}$$

where $X_{core}$, and $(1 - X_{core})$ are core and shell volume fractions, respectively.

FIG. 2a and 2b show real part of the effective dielectric function of spherical hollow/Au nanoshells with a diameter of the 40 nm for different shell thicknesses without and with electron-surface collision effect, respectively. It is evident from FIG. 2 and FIG. S2 that electron-surface collision has little to no effect on the real part of the electric permittivity. However, as can be understood from these figures, the slope and amount of $\varepsilon_{1eff}$ significantly alter with changing the shell thickness. Since the dielectric function of the air ($\varepsilon_{air} = 1$) is dispersionless, and $\varepsilon_1$ of Au is negative and varying with the wavelength of incident light, the core volume fraction will increase that results in increase of $\varepsilon_{1eff}$, and reducing the slope with a decrease in shell thickness. On the other hand, $\varepsilon_{2eff}$ depends on two factors: volume fraction of the shell and electron-surface scattering rate. As shown in FIG. 2c and 2d, the $\varepsilon_{2eff}$ decreases by reducing shell thickness. However, it has already been noted above that by decreasing effective size of the shell electron-surface collision frequency increases. As demonstrated in FIG. S2b, the electron-surface effect will become important for longer wavelengths and leads to increasing $\varepsilon_{2eff}$ with reducing shell thickness. In hollow nanoshell structures, reducing shell thickness increases the real part of the dielectric function while decreases its gradient (See FIG. 2a and 2b). To satisfy Eq. 6 for plasmon resonance, with a decreasing gradient of the real part of the dielectric function, it requires the LSPR peak to shift significantly with changing the refractive index of the medium. Thus, decreasing shell thickness for fixed total size leads to increased sensitivity for hollow-Au nanoshells.

To characterize the LSPR sensitivity, the FWHM of the LSPR resonance spectrum has been obtained by Taylor expansion of the absorption spectrum at LSPR wavelength as[25,79]

$$\Delta\lambda_{1/2} \approx 2\left|\lambda_R - \lambda_{1/2}\right| \approx 2\varepsilon_2(\lambda_{LSPR})/\frac{d\varepsilon_1(\omega)}{d\lambda}\Big|_{\lambda=\lambda_{LSPR}} \tag{9}$$

FWHM plays an important role to determine relative scattering and absorption to extinction. Eq. 9 suggests that in order to have narrower bandwidth it is required to optimize NP structures to have smaller imaginary part of the dielectric function and steeper derivative of the real part at the resonance frequency. As it is demonstrated in FIG. 2 for spherical hollow-Au nanoshells with a total diameter of 40 nm, both the imaginary part and slope of the real part of the dielectric function reduce by decreasing shell thickness. From this result, it can be concluded that the imaginary part of dielectric functions and gradient of the real part of structures play important role to obtain higher sensitivity and lower FWHM.

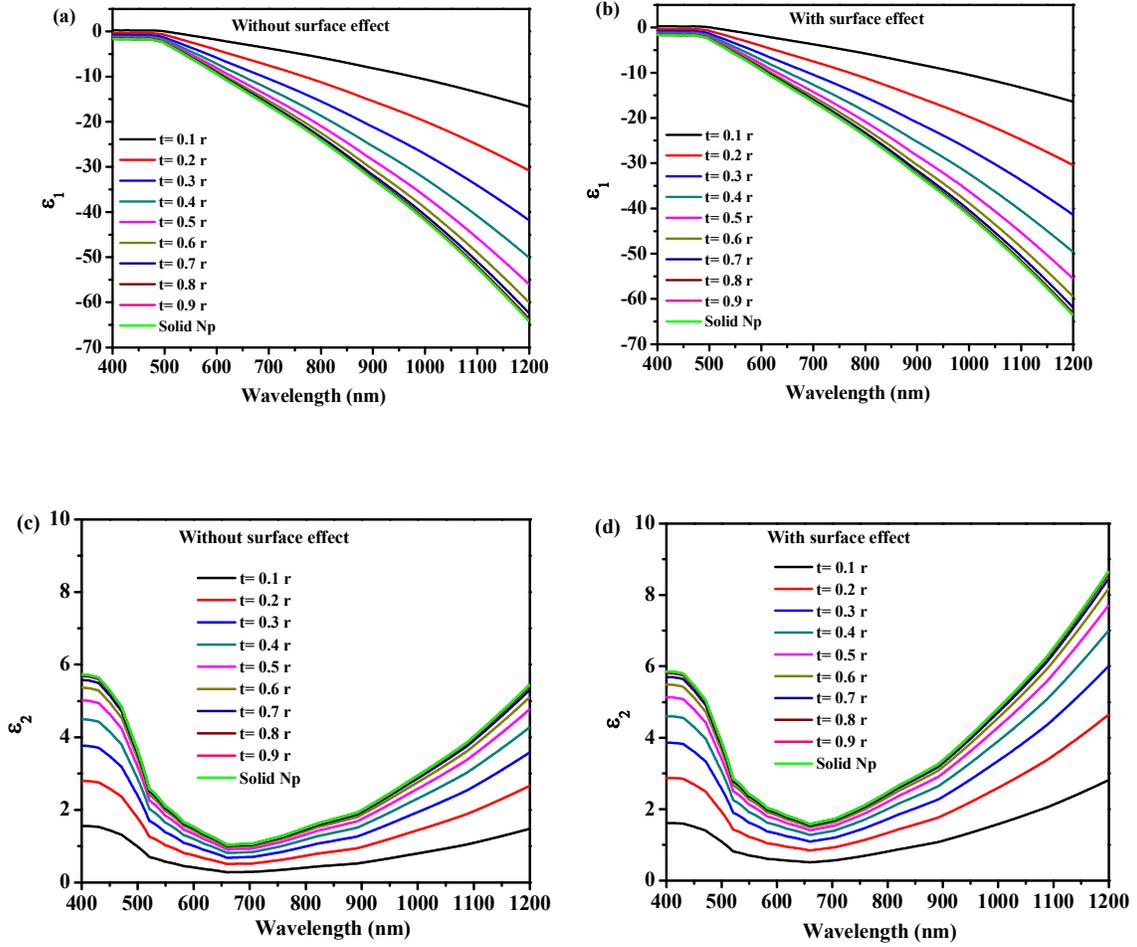

**FIG. 2.** Changing real ($\varepsilon_1$) and imaginary part ($\varepsilon_2$) of the dielectric function of hollow Au nanoshell with a total diameter of 40 nm for different shell thicknesses versus incident light wavelength: (a) and (c) without considering electron-surface scattering effect, (b) and (d) with considering electron-surface scattering effect in damping frequency. For comparison, electric permittivity of solid NPs in both cases has been added.

## B. Size effect on the LSPR sensitivity

In this section, the size effect on the LSPR sensitivity of spherical hollow-Au nanoshells with varying shell thickness was studied. Two layers Mie theory was used to calculate the extinction spectrum and corresponding LSPR wavelength of each sample (See FIG. S3). FIG. 3a-3d show LSPR peak wavelength versus refractive index of the medium for different shell thicknesses for nanoparticles with a total diameter of d=20 nm, d=40 nm, d=60 nm, and d=80 nm, respectively. As shown in FIG. 3, $\lambda_{LSPR}$ and its corresponding gradient significantly increases with reducing the ratio of shell thickness-to-total radius (t/r). The two possible explanations for this result are: 1) enhancing inner and outer layers plasmonic coupling due to reducing the shell thickness,[64,80,81] 2) decreasing gradient of the real part of the dielectric function ($\frac{d\varepsilon_1(\omega)}{d\lambda}$).

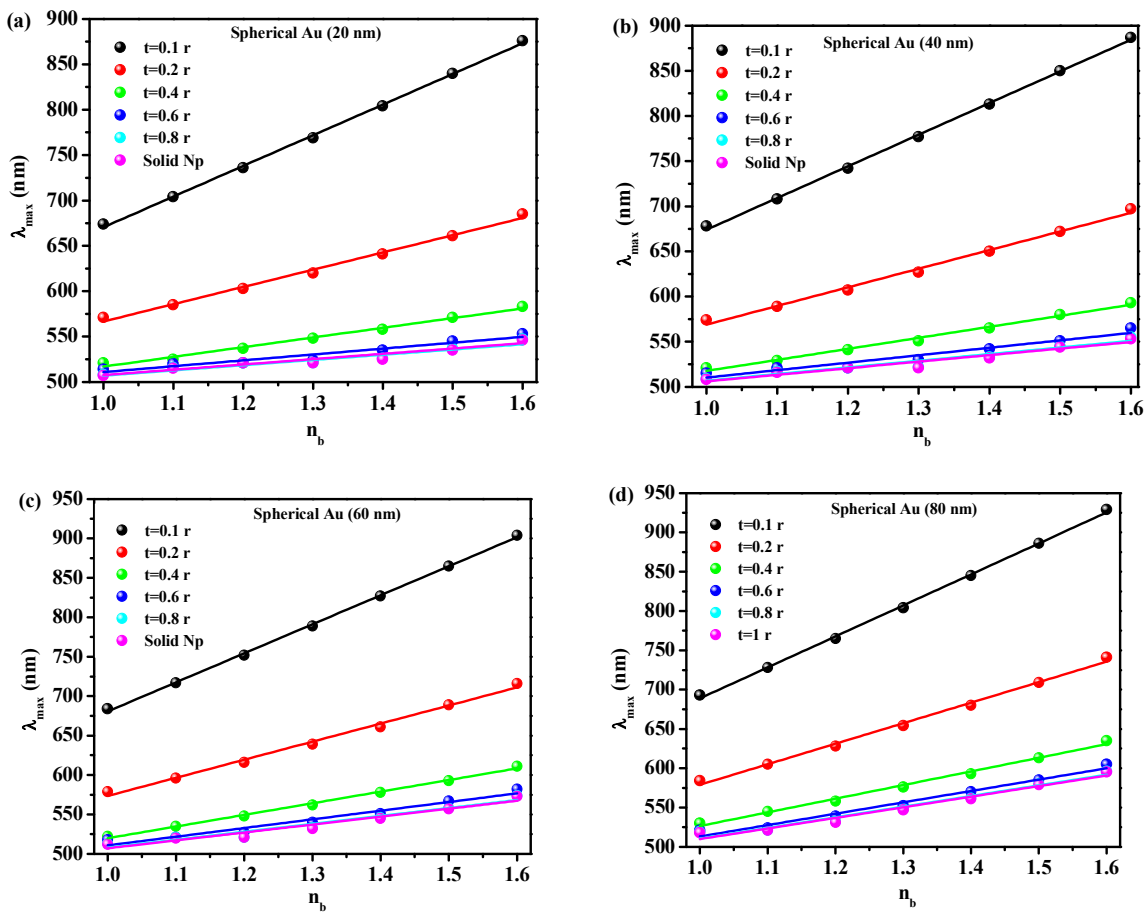

**FIG. 3.** LSPR peak wavelength of the spherical hollow/Au nanoshells for different shell thicknesses versus RI of the surrounding medium with the diameter of (a) d=20 nm (b) d=40 nm, (c) d=60 nm and (d) d= 80 nm.

The LSPR sensitivity is obtained by the slope of the linear fit of the LSPR wavelength versus RI of the surrounding medium and is presented in FIG. 4. Similar to previous studies[53,64], our finding shows that the LSPR sensitivity increases with reducing shell thickness, at an exponential rate. As an example, for Np with d=20 nm, the LSPR sensitivity increases significantly from 57.5 to 338 nm/RIU by reducing shell thickness from t=1 r (Solid Sphere) to t=0.1 r. In addition, the sensitivity increases with increasing the size due to increasing contribution from higher order modes.[25,82] However, by reducing the shell thickness, the LSPR sensitivity of the different size NPs converge. This can be ascribed to the decrease in contribution of the higher order modes leading to size independent behavior of LSPR sensitivity for thin shells.

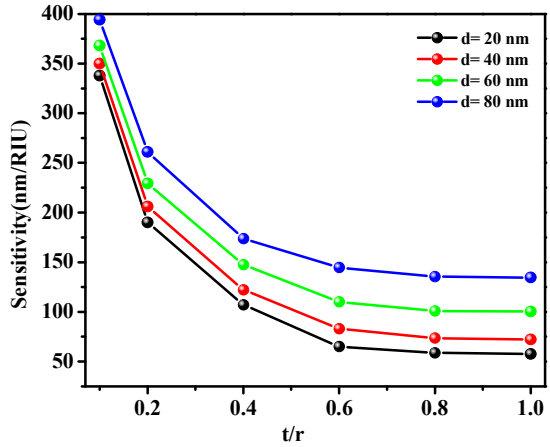 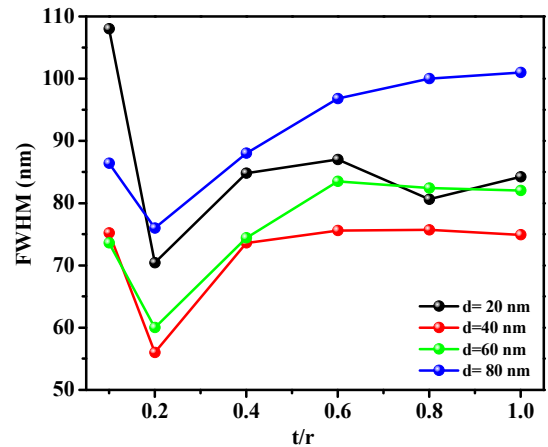

**FIG. 4.** Size effect on the LSPR sensitivity of the spherical hollow/Au nanoshells for the NPs with the total diameter of d=20 nm, d=40 nm, d=60 nm, and d= 80 nm.

**FIG. 5.** Evolution of FWHM of spherical hollow/Au nanoshells versus shell thickness-to-total radius (t/r) for the NPs with the total diameter of d=20 nm, d= 40 nm, d=60 nm and d= 80 nm.

In order to easily detect the LSPR peak shift with changing RI of the surrounding medium, structures with lower FWHM that have higher extinction (See FIG. S5) are desirable.[25] FWHM in nanoshells depends on several factors: electron- surface scattering at the boundaries of the nanoshell; shell to core volume ratio; and retardation effect, including the contribution of higher order plasmon modes.[25,82] FIG. 5 shows the calculated FWHM of the different size spherical nanoshells that were obtained using Gaussian fit to extinction spectrum of the NPs in aqueous solution ($n_m = 1.34$). Interestingly, the plasmon bandwidth is minimum at t=0.2 r for all sizes. This observation could be attributed to the fact that the reduction in imaginary part of the dielectric function is dominated by a decrease of $\frac{d\varepsilon_1(\omega)}{d\lambda}$ leading to a decrease of the FWHM of the nanoshells. Interestingly, the results also show that with an increase in the size of NPs from 20 nm to 40 nm, FWHM decreases and then increases by further enlarging the NPs size. The inconsistency may be due to the increased retardation effect and contribution of the higher plasmonic modes that overcome the reduced electron-surface scattering rate effect with increasing size of the NP from 40 to 80 nm, and results in a net increase of the plasmon bandwidth.

**C. Shape effect on LSPR sensitivity**

In order to explore the shape effect on the LSPR peak wavelength, and LSPR sensitivity, we have performed DDA simulation for five different shapes (disk, prism, rod, ellipsoid, and rectangular block) with an effective diameter of 40 nm and compared their optical properties with same sized spherical hollow/Au nanoshells. For the rod, ellipsoid and rectangular block, the aspect ratio was set to 2 and direction of the incident light electric field has been considered parallel to the long axis of the nanoshells. For the prism, the thickness was considered to be half of the side length of equilateral triangular and incident light has been chosen in the direction of the trigonal axis. The thickness of the disk nanoshell was set to be equal to its radius, and the electric field was chosen in the direction of its main axis. DDA calculation was performed for varying shell thickness of each sample while the total effective diameter of each was fixed to 40 nm. FIG. S6 shows the evolution of the LSPR peak wavelength of each sample with changing RI of the medium for different shell thicknesses. As a reference, the calculated LSPR of the spherical hollow/Au nanoshells with a total diameter of 40 nm was added to the figure. It is observed that LSPR shift for a given surrounding medium RI increases by reducing shell thickness in all samples. As shown in FIG. S6 and FIG. S.7, for a given RI change of the surrounding medium, the LSPR peak of the ellipsoid, rod and rectangular block nanoshells shifted to longer wavelengths more than other shaped shells and can be tuned from visible to infrared region by changing the shell thickness. These results are mainly due to lower depolarization factor or higher polarizability of the ellipsoid, rod and rectangular block nanoshells when compared to other structures.[5] By decreasing the depolarization factor, the restoring Coulombic force between the positive ions and oscillating electrons decreases which decreases the required incident light energy to excite the free electrons of the nanoshell.[5]

The results show that the sensitivity of the rectangular block and the disk is the highest and lowest for all shell thicknesses, respectively (see FIG. 6). For example, at t=0.1 $r_{eff}$, the LSPR sensitivity of the rectangular block is 594.9 nm/RIU which is almost 1.7 times of the sphere (350 nm/RIU) and 2.1 times of the disk (278.4 nm/RIU) with same shell thickness. In addition, LSPR sensitivity of rod and ellipsoid converge to rectangular block by further reducing the shell thickness. The sensitivity of rod nanoshells is slightly higher than ellipsoid in all shell thickness, which can be ascribed to the larger average distance between positive ions and excited electrons in rod nanoshell.

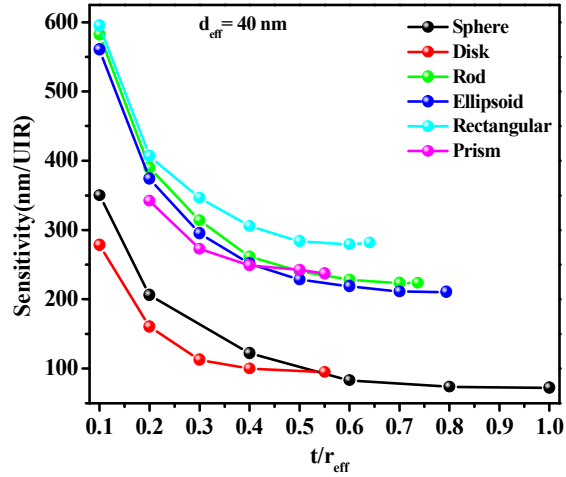

**FIG. 6.** Influence of shape and shell thickness on the LSPR sensitivity of the sphere, disk, rod, ellipsoid, rectangular block, and prism, shaped hollow/Au nanoshell with an effective diameter of $d_{eff}$ = 40 nm.

FIG. 7 shows the summary of the FWHM of the studied samples with varying shell thickness. It shows a clear trend of exponential decreasing of the FWHM for ellipsoid, rod, rectangular block, and prism with increasing shell thickness. The significant reduction has happened when the shell thickness varied from t=0.1 $r_{eff}$ to t=0.2 $r_{eff}$. At t=0.2 $r_{eff}$, the FWHM reaches 86.8, 75.62, 73.1 and 74.3 nm for ellipsoid, rod, rectangular block, and prism, respectively. However, the FWHM of spherical and disk reduces just in the range of t=0.1- 0.2 $r_{eff}$ and reach its minimum value at t= 0.2 $r_{eff}$, and then increases with further increasing the shell thickness. In the next section, the aspect ratio effect on the optical properties of the rectangular block and rod hollow/Au nanoshells is presented.

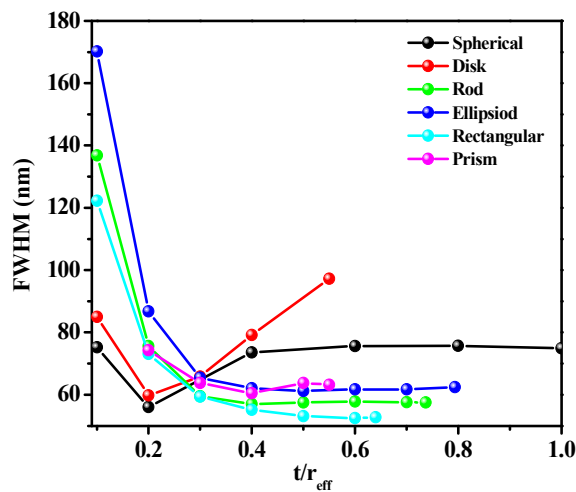

**FIG.7.** Influence of shape and shell thickness on the FWHM of the sphere, disk, rod, ellipsoid, rectangular block, and prism, shaped hollow/Au nanoshell with an effective diameter of the $d_{eff}$ = 40 .

### D. Aspect ratio effect on the sensitivity

Increasing AR of the NPs will enhance surface area-to-volume ratio that can make sensing of target molecules easy by providing enhanced active area.[33] To study AR effects in sensitivity, we investigated optical properties of the rectangular block and rod nanoshells which have shown higher sensitivity for AR=2 when compared to other structures, as discussed in the previous section. The effective total diameter of the samples was set to 40 nm, and the AR was varied from 1 to 5. Note that at AR=1, Rod and Rectangular block will convert to spherical and cube, respectively. The DDA calculation was performed for three different cases: 1) Solid Np, 2) nanoshell with fixed t=0.2 $r_{eff}$ and 3) nanoshell with fixed shell volume. In the third case, the volume of the shell for all aspect ratios was set to be equal to the volume of the shell with AR=2 and t=0.2 $r_{eff}$.

FIG. 8a and 8b show the LSPR sensitivity of the rod and the rectangular block versus AR, respectively. It is observed that sensitivity for all cases increases with increasing AR. This can again be ascribed to enhancing polarizability or decreasing depolarization factor. Also, it was demonstrated that the sensitivity for both hollow/Au nanoshells is significantly higher than the solid NPs due to strong plasmon coupling between the inner and outer layer of the shell. However, it must be noted that the sensitivity of the structures with fixed shell volume diverges and become higher than the case with fixed shell thickness with an AR greater than 2. This observation is due to the fact that in the case with fixed shell volume, shell thickness slightly reduces with increase in AR. For example, at an AR of 5 and for fixed shell volume, the LSPR sensitivity of rectangular block is 800 nm/RIU that is ,despite its smaller size, almost same as the longitudinal plasmons mode of the nanorice (801.4 nm/RIU) with longitudinal diameter of 340± 20 nm, transverse diameter of 54± 4 nm and shell thickness of 13.1± 1.1 nm.[53]

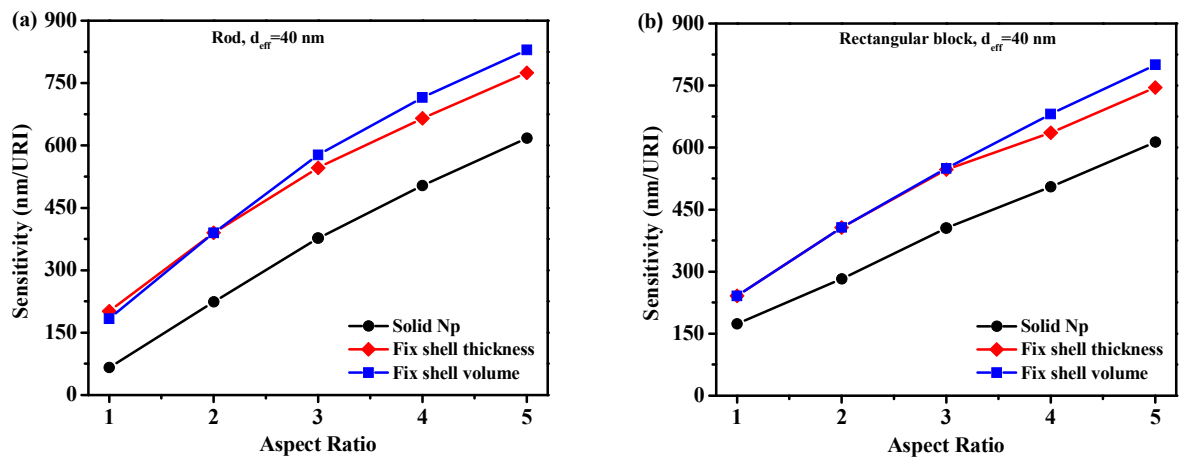

**FIG. 8.** The sensitivity of the hollow/Au nanoshells, (a) Rod and (b) Rectangular block, with $d_{eff}$ = 40 nm versus aspect ratio

A comparison of the two results reveals that for small AR (<3), the sensitivity of the rectangular block is more than the rod, whereas by further increase in AR, the rod sensitivity exceeds that of rectangular block. The interesting finding was that the LSPR sensitivity of the nanorod with an AR of 5 and for fixed shell volume is 829.5 nm/RIU which is higher than the structures mentioned above (rectangular block and nanorice). However, it must be noted that further increase of the AR reduces the extinction spectra due to increasing electron-surface scattering rate, which makes detection of the wavelength changes experimentally difficult. As shown in FIG. 9, by increasing the AR of both rod and rectangular block the extinction spectra will increase, while the amount of enhancement reduces by increasing the AR.

In summary, it has been shown that both the rectangular block and rod hollow/Au nanoshells provide higher sensitivity in comparison to other structures. Also, due to their smaller sizes, the RI of the surrounding medium can change effectively with a low binding concentration of the target molecules, which offers the possibility to sense and detect the analytes such as DNA which have low concentrations. The tunability of the sensitivity of these structures with changing the shell thickness and AR provides the opportunity to further enhance LSPR sensitivity and molecules detecting capabilities of the single NPs.

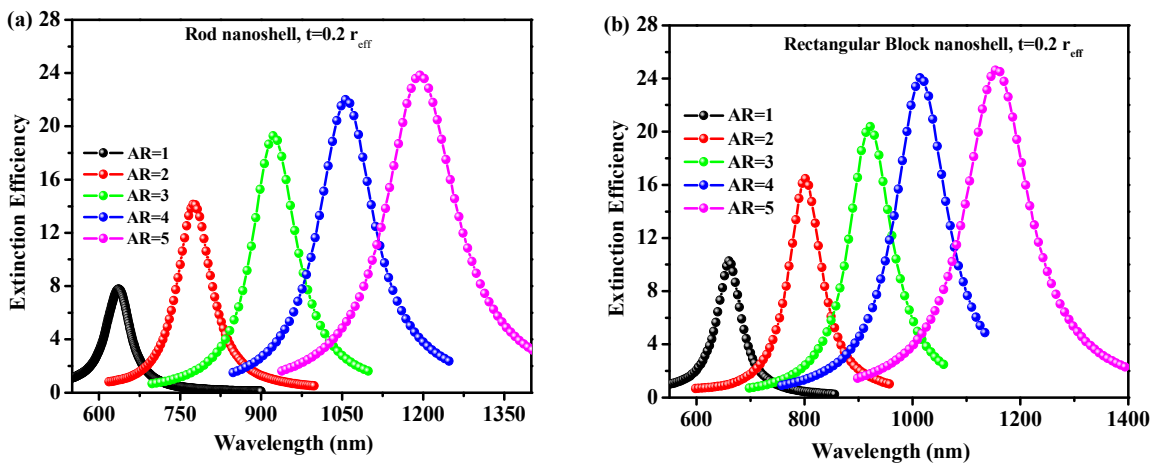

FIG. 9. Extinction efficiency spectra of the (a) rod shape, and b) rectangular block, hollow/Au nanoshells with an effective diameter of 40 nm, and shell thickness of t=0.2 $r_{eff,}$ for varying the AR from 1 to 5.

## IV. CONCLUSIONS

The effect of the size, shape, shell thickness and AR on the LSPR sensitivity of hollow-Au nanoshell structures was studied. Six different shaped hollow/Au nanoshells structures were investigated that include: sphere, disk, rod, ellipsoid, rectangular block, and prism. From the results, it was observed that by reducing the shell thickness of the shells, the plasmonic coupling between the inner and outer surface of the Au shell increases that leads to shift of LSPR maximum to longer wavelengths and increasing its sensitivity. Also, results show that LSPR sensitivity increases with an increase in the size of NPs. From shape effect studies, it was demonstrated that rod and rectangular block nanoshells show higher sensitivity when compared to other samples. In addition, the AR effect was simulated for rod and rectangular block nanoshells, which show higher sensitivity when compared to other structures. It is found that for both structures, LSPR peak shifts more and its sensitivity enhanced by increasing the AR of the nanoshells. This study has found that rod and rectangular block hollow/Au nanoshells possess higher LSPR sensitivity in comparison to other structures, which make them suitable candidates for sensing applications.

## SUPPLEMENTARY MATERIAL

**Supporting Information Available**

## ACKNOWLEDGEMENTS
We thank physics department of Western Michigan University for financial support.

# Supporting Information

# Theoretical Investigation of Size, Shape, and Aspect Ratio effect on the Sensitivity and Figure of Merit of Hollow-Gold Nanoshells

M. Shabaninezhad[a], and R. Guda[b]

[a]Department of Physics, Western Michigan University, Kalamazoo MI 49008

[b]Department of Chemistry, Western Michigan University, Kalamazoo MI 49008

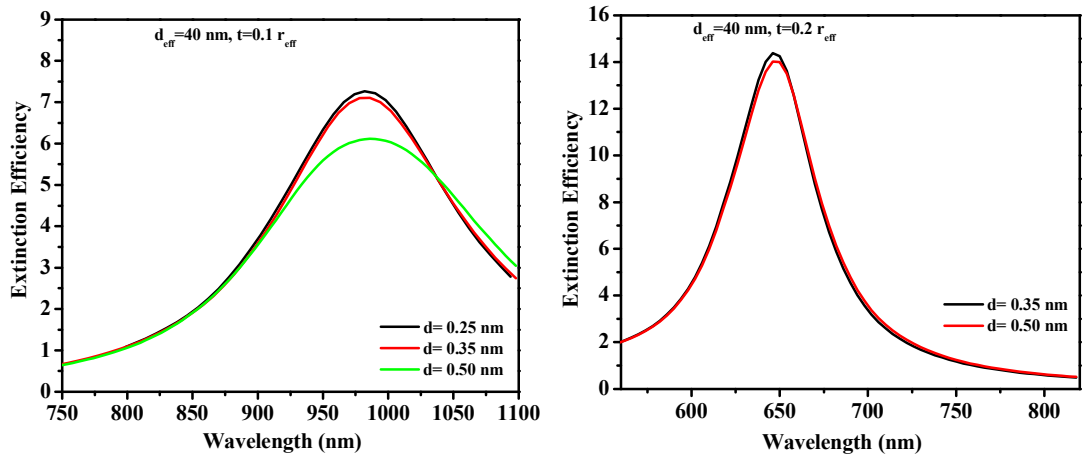

**FIG. S. 1.** Convergence of the extinction spectra of the ellipsoid hollow/Au nanoshells with an effective diameter of 40 nm and a shell thickness of a) t=0.1 reff and b) t=0.2 reff. As shown in these figures, for t=0.1 $r_{eff}$ and t=0.2 $r_{eff}$ extinction spectra converge at d=0.35 nm and d=0.5 nm, respectively. The incident light electric field is parallel to the main axis of the NPs.

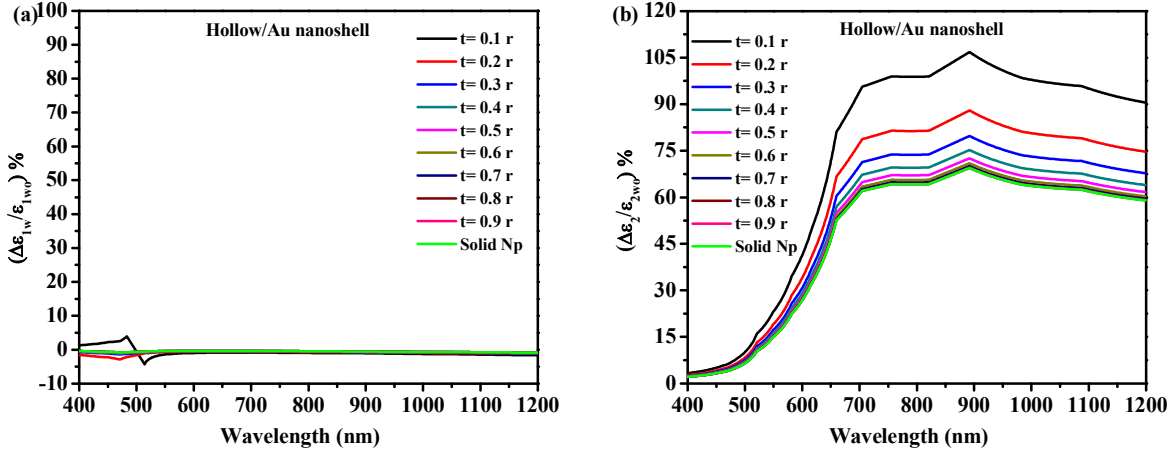

**FIG. S. 2.** Percentage of changing of the (a) real part and (b) the imaginary part, of the dielectric function of spherical hollow/Au nanoshells with a total size of d=40 nm with varying shell thickness due to considering surface effect ($\Delta\varepsilon = \varepsilon_w - \varepsilon_{wo}$, where $\varepsilon_w$ and $\varepsilon_{wo}$ are the dielectric function of the hollow/Au nanoshell with and without considering the surface effect, respectively). Here $\varepsilon_{1wo}$ and $\varepsilon_{2wo}$ are real and imaginary part of the dielectric function of hollow/Au nanoshells without electron- surface scattering effect.

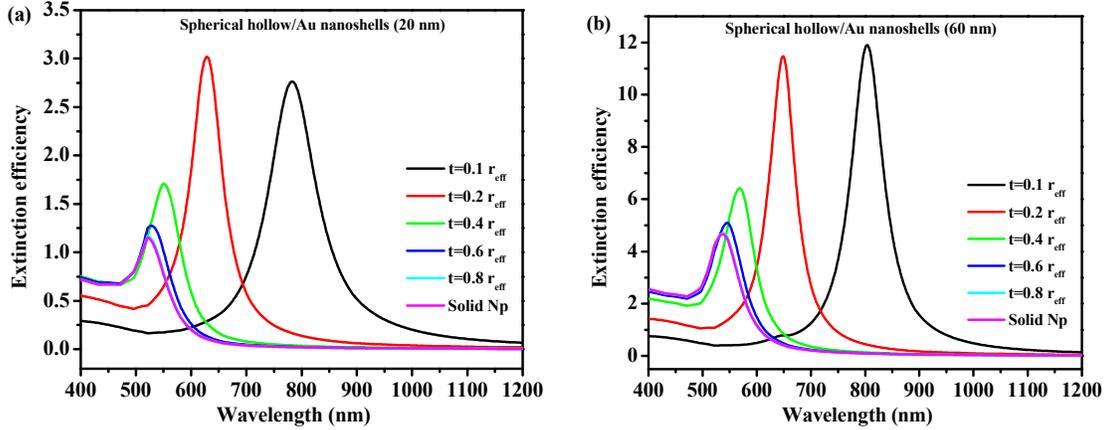

**FIG. S.3.** Extinction spectrum of spherical hollow/Au nanoshells for different shell thickness with a total diameter of (a) 20 nm and (b) 60 nm.

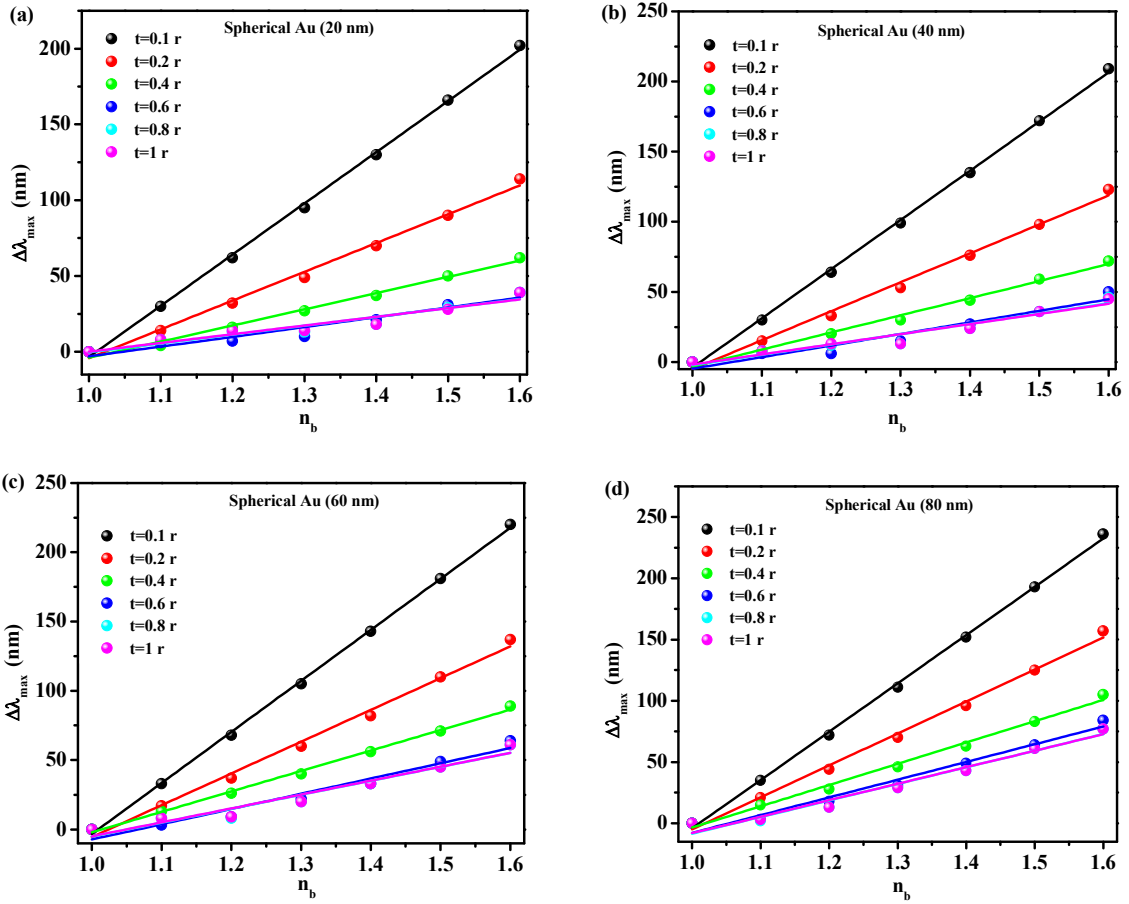

**FIG. S. 4.** LSPR peak wavelength shift of the spherical hollow/Au nanoshells with varying shell thicknesses for the diameter of (a) d=20 nm, (b) d=40 nm (c) d=60 nm, and (d) d= 80 nm, versus the RI of the surrounding medium

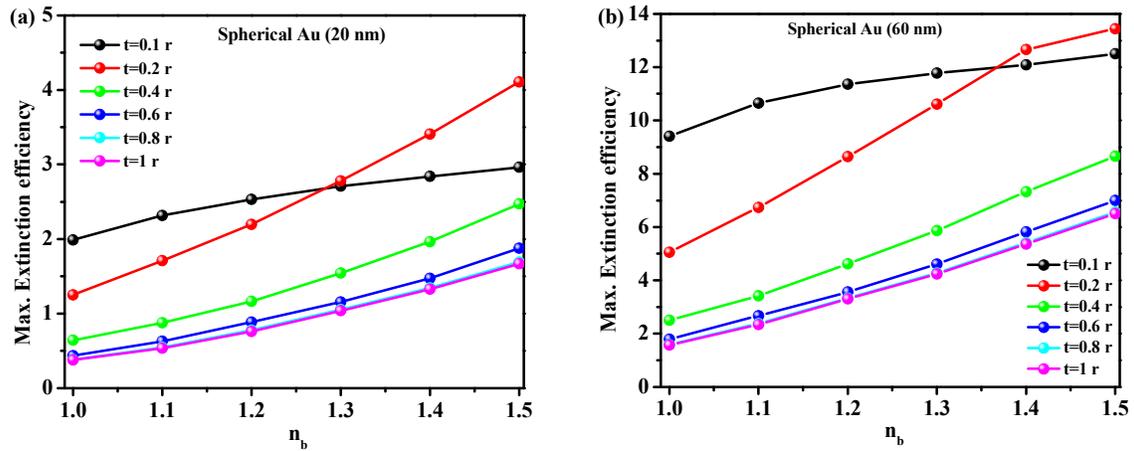

**FIG. S. 5.** Maximum extinction efficiency of the spherical hollow/Au nanoshells against RI of the surrounding medium for different shell thickness with a total diameter of (a) 20 nm and (b) 60 nm.

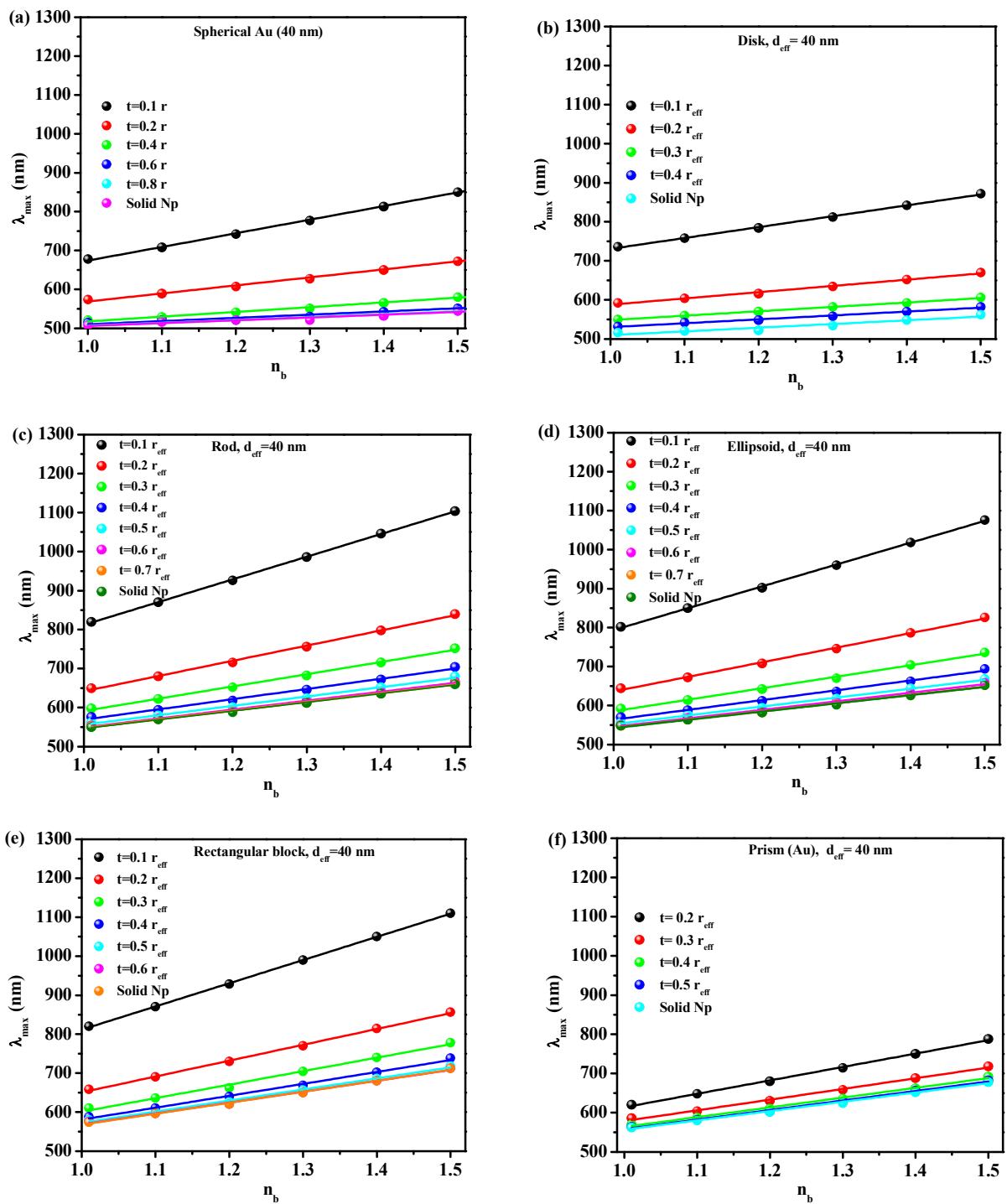

**FIG. S. 6.** LSPR peak wavelength of the hollow/Au nanoshells versus RI of the surrounding medium for different shapes: (a) Spherical, (b) Disk, (c) Rod, (d) Ellipsoid, (e) Rectangular block and (f) Prism, with $d_{eff}$ = 40 nm. The incident light electric field was chosen to be parallel to the main axis of the NPs.

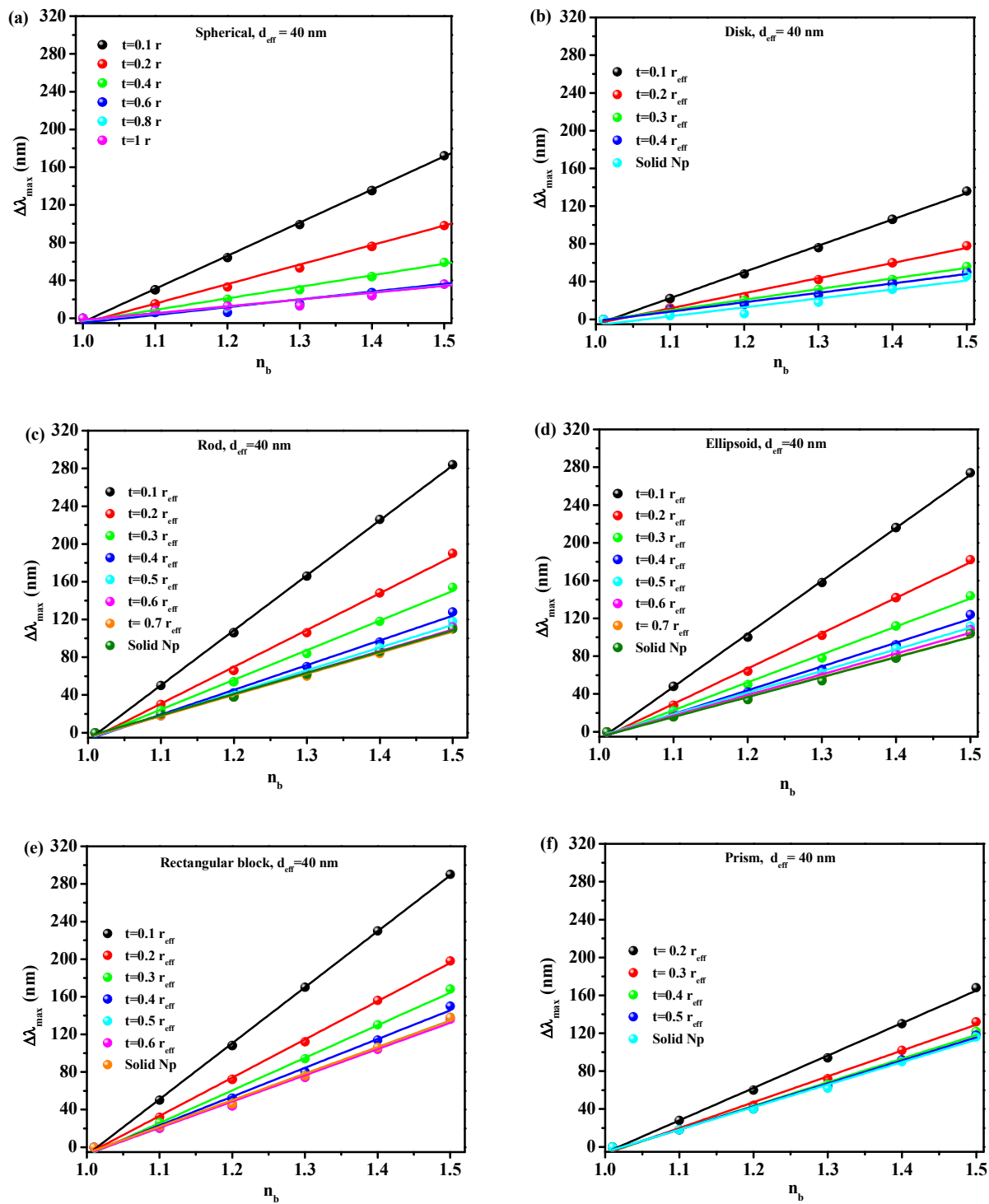

**FIG. S. 7.** LSPR peak wavelength shift of the hollow Au nanoshells versus RI of the surrounding medium for different shapes: (a) Spherical, (b) Disk, (c) Rod, (d) Ellipsoid, (e) Rectangular block and (f) Prism, with $d_{eff}$=40 nm. The incident light electric field was chosen to be parallel to the main axis of the NPs.

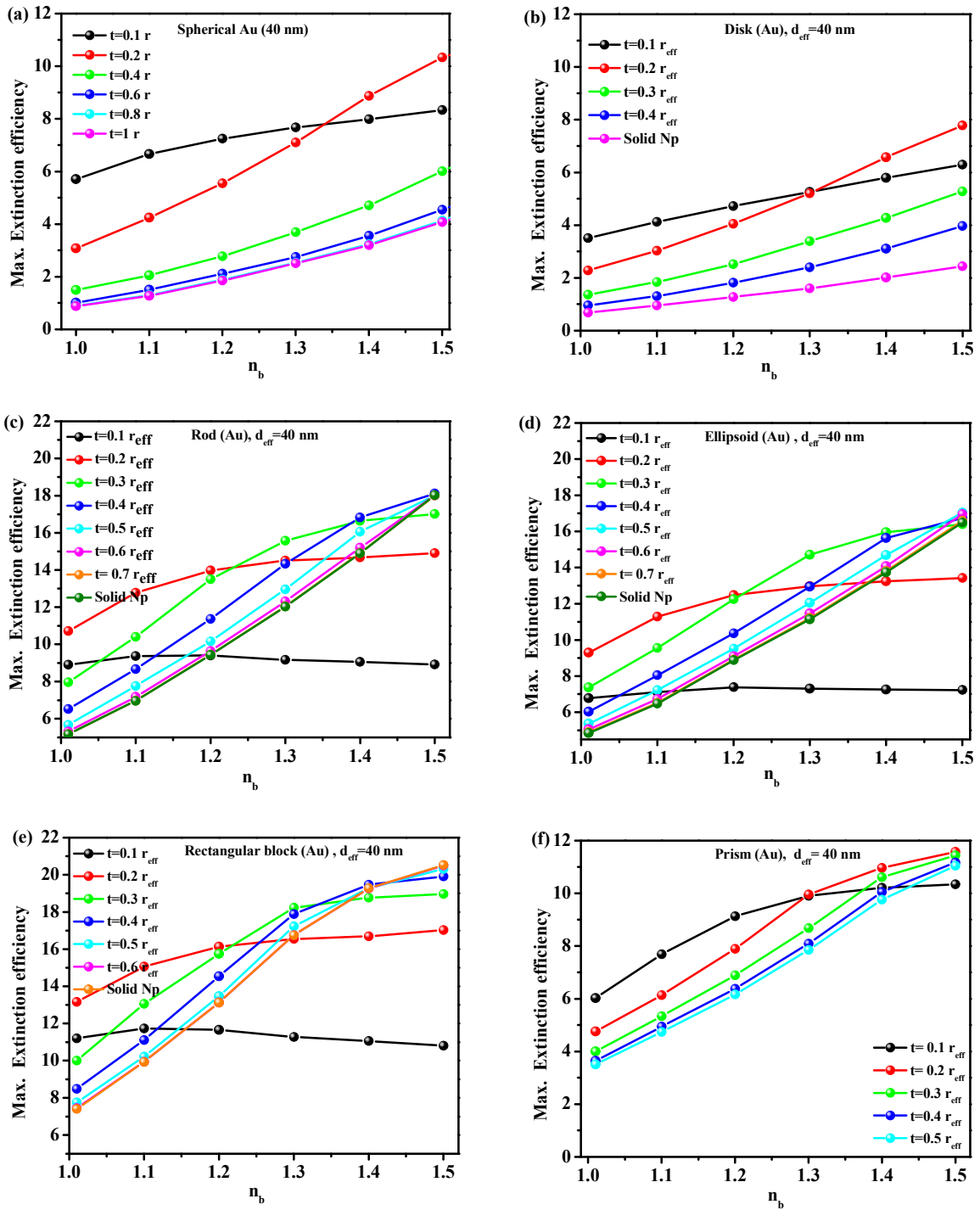

**FIG. S.8.** Maximum extinction efficiency of spherical hollow Au nanoshells versus RI of the surrounding medium for different shapes: (a) Spherical, (b) Disk, (c) Rod, (d) Ellipsoid, (e) Rectangular block and (f) Prism, with deff = 40 nm. The incident light electric field is parallel to the main axis of the NPs.

25